\documentclass[twocolumn,showpacs,preprintnumbers]{revtex4}%
\usepackage{amsmath}
\usepackage{amssymb}
\usepackage{graphicx}
\usepackage{dcolumn}
\usepackage{bm}
\usepackage{amsfonts}%
\begin{document}
\preprint{ }
\title{Performance of cavity-parametric amplifiers, employing Kerr nonlinearites, in
the presence of two-photon loss}
\author{Bernard Yurke}
\affiliation{Bell Laboratories, Lucent Technologies, 600 Mountain Avenue, Murray Hill, NJ 07974}
\author{Eyal Buks}
\affiliation{Department of Electrical Engineering, Technion, Haifa 32000 Israel}
\date{\today }

\begin{abstract}
Two-photon loss mechanisms often accompany a Kerr nonlinearity. The kinetic
inductance exhibited by superconducting transmission lines provides an example
of a Kerr-like nonlinearity that is accompanied by a nonlinear resistance of
the two-photon absorptive type. Such nonlinear dissipation can degrade the
performance of amplifiers and mixers employing a Kerr-like nonlinearity as the
gain or mixing medium. As an aid for parametric amplifier design, we provide a
quantum analysis of a cavity parametric amplifier employing a Kerr
nonlinearity that is accompanied by a two-photon absorptive loss. Because of
their usefulness in diagnostics, we obtain expressions for the pump amplitude
within the cavity, the reflection coefficient for the pump amplitude reflected
off of the cavity, the parametric gain, and the intermodulation gain.
Expressions by which of the degree of squeezing can be computed are also presented.

\end{abstract}
\maketitle





\section{Introduction}

Sensitive superconducting microwave devices such as SIS mixers
\cite{tucker79,tucker85} and parametric amplifiers \cite{kuzmin83,yurke89}
have been devised which achieve performances close to the quantum limit.
Phase-sensitive Josephson-junction parametric amplifiers have been constructed
whose noise performance exceeds that of the quantum limits imposed on linear
phase-insensitive parametric amplifiers \cite{movshovich90}. These
phase-sensitive amplifiers have been used to generate quantum mechanical
states of the electromagnetic field, called squeezed states, whose noise in
one amplitude component is reduced below that of vacuum fluctuations. The
kinetic inductance of superconducting transmission lines could also be used to
make low-noise parametric amplifiers. However, associated with the kinetic
inductance is a nonlinear resistance that can degrade device performance.
These nonlinear effects are relatively strong in superconducting striplines
and microstrips due to the nonuniform distribution of the microwave current
along the cross section of the transmission line. Along the edges, where the
current density obtains its peak value, the current density can become
overcritical even with relatively moderate power levels. As a result, the
superconducting current density may vary and, consequently, both inductance
$L$ and resistance $R$ per unit length become current dependent according to
the form \cite{Dahm97}%

\begin{equation}
L=L_{0}+\Delta L\left(  \frac{I}{I_{c}}\right)  ^{2} \label{L}%
\end{equation}

\begin{equation}
R=R_{0}+\Delta R\left(  \frac{I}{I_{c}}\right)  ^{2}, \label{R}%
\end{equation}

where $I$ ($I_{c}$) is the total (critical) current. The kinetic inductance
provides a Kerr-like nonlinearity suitable for the construction of parametric
amplifiers which employ four-wave mixing. The nonlinear resistance, to lowest
order, is of the two-photon absorptive type. To aid in the design of
parametric microwave amplifiers which employ kinetic inductance we have
preformed an analysis of cavity parametric amplifiers employing a Kerr
nonlinear element for gain and a two-photon absorptive loss. Although the
analysis was carried out with a specific application in mind \cite{abdo}, it
is more generally applicable, since two-photon absorptive processes often
accompany Kerr nonlinearities. There are optical
\cite{villeneuve93,fox95,ho95} and mechanical \cite{zaitsev} systems with such
combinations of nonlinearities.

Squeezing in a parametric amplifier with a two-photon absorber has been
studied by a number of workers \cite{gerry93,gilles94,li95,ho95}. In the
analysis provided here, we present expressions for the amplitude of the pump
field within the cavity, the reflection coefficient for the pump off the
cavity, the intermodulation gain, and the degree of squeezing. The first,
second, and third of these quantities are particularly useful for extracting
model parameters from experimental data. The equations of motion are derived
using the input-output theory of Gardiner and Collett \cite{gardiner85,Gea90}.
The undepleted pump approximation is then made, allowing the pump field inside
the cavity and the pump field reflected from the cavity to be calculated. The
small signal response is then obtained by linearization about the pump field.

\section{The Hamiltonian}

A lossless transmission line resonator having nonlinear kinetic inductance is
discussed in appendix A and the effect of nonlinear losses associated with the
kinetic inductance is discussed in appendix B. \ Here we consider the case
where the external signals employed for externally driving the resonator are
all close in frequency to one of the resonances at $\omega_{0}$. As we discuss
in the appendix, under some conditions, which are assumed to be satisfied, all
other modes of the resonator can be disregarded. In this case the Hamiltonian
of the nonlinear resonator can be written as \cite{imoto85,white00}%

\begin{equation}
H_{r}=\hbar\omega_{0}A^{\dagger}A+\frac{\hbar}{2}KA^{\dagger}A^{\dagger}AA,
\label{H_r}%
\end{equation}

where the Kerr constant $K$ is given in Eq. \ref{K_n'}.

As seen in Fig. 1, the resonator is coupled to a test port (labeled as $a_{1}%
$) serving as the input-output port. Operated as an amplifier, the signal
returned or "reflected" from the input port is larger than the incoming
signal. This mode of operation, at microwave frequencies, is referred to as
the negative-resistance reflection mode. Two extra fictitious ports are added
in order to theoretically model dissipation \cite{yurke84}. Port $a_{2}$
serves as a linear loss port. Port $a_{3}$ serves as the two-photon loss port.
The coupling of the $a_{3}$ loss mode to the resonator mode $A$ is nonlinear
and given in Eq.~(\ref{eq:h9}).%

\begin{figure}
[ptb]
\begin{center}
\includegraphics[
trim=1.075247in 4.301081in 1.076348in 1.434260in,
height=1.1865in,
width=3.7421in
]%
{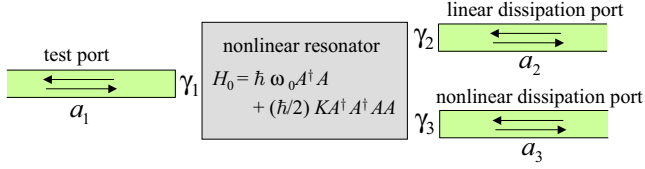}%
\caption{ The model includes nonlinear resonator coupled to three ports, a
test port, a linear dissipation port and a non-linear one.}%
\end{center}
\end{figure}

It is convenient to write the Hamiltonian as a sum of terms,
\begin{equation}
H=H_{r}+H_{a_{1}}+H_{a_{2}}+H_{a_{3}}+H_{T_{1}}+H_{T_{2}}+H_{T_{3}}\ ,
\label{eq:h1}%
\end{equation}
each representing the Hamiltonian for a component of the system.

The three ports coupled to the resonator (see Fig. 1) serve as baths. One bath
models the external modes that couple to the resonator mode through the port
that serves both as the input port and as the output port. The Hamiltonian
$H_{a1}$ for this bath is given by
\begin{equation}
H_{a1}=\int d\omega\hbar\omega a_{1}^{\dagger}(\omega)a_{1}(\omega)\ .
\label{eq:h4}%
\end{equation}
The other two baths are associated with the linear and nonlinear cavity losses
and their Hamiltonians are given by
\begin{equation}
H_{a2}=\int d\omega\hbar\omega a_{2}^{\dagger}(\omega)a_{2}(\omega)
\label{eq:h5}%
\end{equation}
and
\begin{equation}
H_{a3}=\int d\omega\hbar\omega a_{3}^{\dagger}(\omega)a_{3}(\omega)\ .
\label{eq:h6}%
\end{equation}
The linear coupling of the bath modes $a_{1}$ and $a_{2}$ to the cavity mode
$A$ is modeled by the hopping Hamiltonians
\begin{equation}
H_{T_{1}}=\hbar\int d\omega\lbrack\kappa_{1}A^{\dagger}a_{1}(\omega
)+\kappa_{1}^{\ast}a_{1}^{\dagger}(\omega)A] \label{eq:h7}%
\end{equation}
and
\begin{equation}
H_{T_{2}}=\hbar\int d\omega\lbrack\kappa_{2}A^{\dagger}a_{2}(\omega
)+\kappa_{2}^{\ast}a_{2}^{\dagger}(\omega)A]\ . \label{eq:h8}%
\end{equation}
The two-photon absorptive coupling of the resonator mode to the bath modes
$a_{3}$ is modeled by a hopping Hamiltonian in which two cavity photons are
destroyed for every bath photon created \cite{tornau74, agarwal86, gilles93,
ezaki99, kitamura99}
\begin{equation}
H_{T_{3}}=\hbar\int d\omega\lbrack\kappa_{3}A^{\dagger}A^{\dagger}a_{3}%
(\omega)+\kappa_{3}^{\ast}a_{3}^{\dagger}(\omega)AA]\ . \label{eq:h9}%
\end{equation}
All the modes in this model satisfy the usual boson commutation relations.

\section{The equations of motion}

Since the creation and annihilation operators appearing in Eqs.~(\ref{eq:h2})
through (\ref{eq:h9}) do not have an explicit time dependence, the Heisenberg
equation of motion for these operators has the form
\begin{equation}
i\hbar\frac{dO}{dt}=[O,H]
\end{equation}
where $H$ is the total Hamiltonian. Using the boson commutation relation for
the cavity mode%
\begin{equation}
\lbrack A,A^{\dagger}]=AA^{\dagger}-A^{\dagger}A=1\ ,
\end{equation}
one has
\begin{align}
\frac{dA}{dt}  &  =-i\omega_{0}A-iKA^{\dagger}AA\\
&  -i\kappa_{1}\int d\omega a_{1}(\omega)-i\kappa_{2}\int d\omega a_{2}%
(\omega)-i2\kappa_{3}\int d\omega A^{\dagger}a_{3}(\omega)\ .\nonumber
\end{align}
Using the boson commutation relations for the bath modes
\begin{align}
\lbrack a_{i}(\omega),a_{j}^{\dagger}(\omega^{\prime})]  &  =\delta
_{i,j}\delta(\omega-\omega^{\prime})\\
\ [a_{i}(\omega),a_{j}(\omega^{\prime})]  &  =0\ ,
\end{align}
one obtains the following equations for the bath modes $a_{1}(\omega)$,
$a_{2}(\omega)$, and $a_{3}(\omega)$:
\begin{equation}
\frac{da_{1}(\omega)}{dt}=-i\omega a_{1}(\omega)-i\kappa_{1}^{\ast}A\ ,
\label{da1/dt}%
\end{equation}%
\begin{equation}
\frac{da_{2}(\omega)}{dt}=-i\omega a_{2}(\omega)-i\kappa_{2}^{\ast}A\ ,
\label{da2/dt}%
\end{equation}
and
\begin{equation}
\frac{da_{3}(\omega)}{dt}=-i\omega a_{3}(\omega)-i\kappa_{3}^{\ast}AA\ .
\label{da3/dt}%
\end{equation}
Using the standard methods of Gardiner and Collett \cite{gardiner85}, these
equations yield the following equation for the cavity mode $A$ driven by the
incoming bath modes $a_{i}^{in}$:
\begin{align}
\frac{dA}{dt}  &  =-i\omega_{0}A-iKA^{\dagger}AA-\gamma A-\gamma_{3}%
A^{\dagger}AA\label{eq:m1}\\
&  \ \ \ -\ i\sqrt{2\gamma_{1}}e^{i\phi_{1}}a_{1}^{in}(t)-i\sqrt{2\gamma_{2}%
}e^{i\phi_{2}}a_{2}^{in}(t)\nonumber\\
&  -i2\sqrt{\gamma_{3}}e^{i\phi_{3}}A^{\dagger}a_{3}^{in}(t)\ .\nonumber
\end{align}
where
\begin{equation}
\gamma=\gamma_{1}+\gamma_{2}%
\end{equation}
and the $\kappa_{i}$, which in general can be complex, have been reexpressed
in terms of the positive real constants $\gamma_{i}$ and the phases $\phi_{i}$
according to
\begin{align}
\kappa_{1}  &  =\sqrt{\frac{\gamma_{1}}{\pi}}e^{i\phi_{1}}\ ,\\
\kappa_{2}  &  =\sqrt{\frac{\gamma_{2}}{\pi}}e^{i\phi_{2}}\ ,\\
\kappa_{3}  &  =\sqrt{\frac{\gamma_{3}}{2\pi}}e^{i\phi_{3}}\ .
\end{align}
Expressions for $\gamma_{2}$ and $\gamma_{3}$ in terms of linear and nonlinear
resistance of the stripline (see Eq. (\ref{R})) are given in Eqs.
(\ref{gamma_2}), (\ref{gamma_3}). In addition, the methods of Gardiner and
Collett \cite{gardiner85} yield the following relations between the outgoing
bath modes $a_{i}^{out}$, the incoming bath modes $a_{i}^{in}$, and the cavity
mode $A$
\begin{equation}
a_{1}^{out}(t)-a_{1}^{in}(t)=-i\sqrt{2\gamma_{1}}e^{-i\phi_{1}}A(t)\ ,
\label{eq:m2}%
\end{equation}%
\begin{equation}
a_{2}^{out}(t)-a_{2}^{in}(t)=-i\sqrt{2\gamma_{2}}e^{-i\phi_{2}}A(t)\ ,
\label{eq:m3}%
\end{equation}%
\begin{equation}
a_{3}^{out}(t)-a_{3}^{in}(t)=-i\sqrt{\gamma_{3}}e^{-i\phi_{3}}A(t)A(t)\ .
\label{eq:m4}%
\end{equation}
In obtaining these equations a Markov approximation \cite{gardiner85} has been
made such that the boson annihilation operators $a_{i}^{in}(t)$ satisfy the
commutation relations
\begin{align}
&  [a_{i}^{in}(t),a_{j}^{in\dagger}(t^{\prime})]=\delta_{i,j}\delta
(t-t^{\prime})\ ,\label{eq:c1}\\
&  [a_{i}^{in}(t),a_{j}^{in}(t^{\prime})]=0\ . \label{eq:c2}%
\end{align}

\section{Response to a classical pump}

Operated as a negative-resistance reflection amplifier, an intense sinusoidal
field, called the pump, is delivered to the input port of the device. Signals
having frequencies to either side of the pump, but lying within the bandwidth
of the device, will be amplified. The linearization procedure is now carried
out in which the signals entering the input port and the noise entering the
loss ports are considered to be small compared to the pump. The first step is
to calculate the classical response of the device to an intense pump in the
absence of signal and noise. The solution is then used to calculate the
linearized response of the device in the presence of signal and noise.

In order to obtain the response of the device to a classical pump in the
absence of signal and noise one sets the incoming noise terms to zero
\begin{align}
a_{2}^{in}  &  =0\ ,\\
a_{3}^{in}  &  =0\ .
\end{align}
The incoming pump is written as
\begin{equation}
a_{1}^{in}=b_{1}^{in}e^{-i(\omega_{p}t+\psi_{1})}%
\end{equation}
where $b_{1}^{in}$ is a real constant, $\omega_{p}$ is the pump frequency, and
$\psi_{1}$ is the pump phase. The outgoing field will also have an oscillatory
time dependence of frequency $\omega_{p}$ and can be written as
\begin{equation}
a_{1}^{out}=b_{1}^{out}e^{-i(\omega_{p}t+\psi_{1})}%
\end{equation}
where $b_{1}^{out}$ may be a complex constant. Writing $A$ as
\begin{equation}
A=Be^{-i(\omega_{p}t+\phi_{B})}%
\end{equation}
where $B$ is a positive real constant, the equations of motion,
Eqs.~(\ref{eq:m1}) and (\ref{eq:m2}), yield
\begin{equation}
\lbrack i(\omega_{0}-\omega_{p})+\gamma]B+(iK+\gamma_{3})B^{3}=-i\sqrt
{2\gamma_{1}}b_{1}^{in}e^{i(\phi_{1}+\phi_{B}-\psi_{1})} \label{eq:p1_}%
\end{equation}
and
\begin{equation}
b_{1}^{out}=b_{1}^{in}-i\sqrt{2\gamma_{1}}Be^{-i(\phi_{1}+\phi_{B}-\psi_{1}%
)}\ . \label{eq:p2}%
\end{equation}
Multiplying each side of the Eq.~(\ref{eq:p1_}) by its complex conjugate and
introducing
\begin{equation}
E=B^{2}\ , \label{eq:6.9}%
\end{equation}
one obtains
\begin{gather}
E^{3}+\frac{2[(\omega_{0}-\omega_{p})K+\gamma\gamma_{3}]}{K^{2}+\gamma_{3}%
^{2}}E^{2}+\frac{(\omega_{0}-\omega_{p})^{2}+\gamma^{2}}{K^{2}+\gamma_{3}^{2}%
}E\nonumber\\
-\frac{2\gamma_{1}}{K^{2}+\gamma_{3}^{2}}(b_{1}^{in})^{2}=0\ . \label{eq:6.10}%
\end{gather}
This cubic equation will have one real solution and two complex solutions or
three real solutions. For the case when two of the solutions are complex, the
real solution is the physical solution. If there are three solutions, two will
be stable and one unstable, and the device will exhibit bistability. Once $E$
and, hence, $B$ have been determined from Eqs.~(\ref{eq:6.10}) and
(\ref{eq:6.9}), the phase $\phi_{B}$ can be determined from Eq.~(\ref{eq:p1_})
and the amplitude of the reflected pump can then be computed from
Eq.~(\ref{eq:p2}). In Fig. 2 (a), (d), and (g) plots of $B$ as a function of
frequency for three different incoming pump amplitudes are shown. The
frequency pulling of the cavity resonance is clearly seen in (d) and (g) as
the incoming pump amplitude $b_{1}^{in}$ is increased. Also plotted in Fig. 2
(b), (e), and (h) is the reflection coefficient $|b_{1}^{out}/b_{1}^{in}|$ for
the reflected pump amplitude as a function of frequency. If no power were
absorbed by the cavity, the reflection coefficient would be unity. One sees a
dip in the reflected power at the cavity resonance. As the incoming pump
amplitude is increased, this absorption feature also shows frequency pulling,
as can be seen in (e) and (h).%

\begin{figure}
[ptb]
\begin{center}
\includegraphics[
trim=0.840758in 0.000000in 1.052041in 0.000000in,
height=2.6308in,
width=3.3399in
]%
{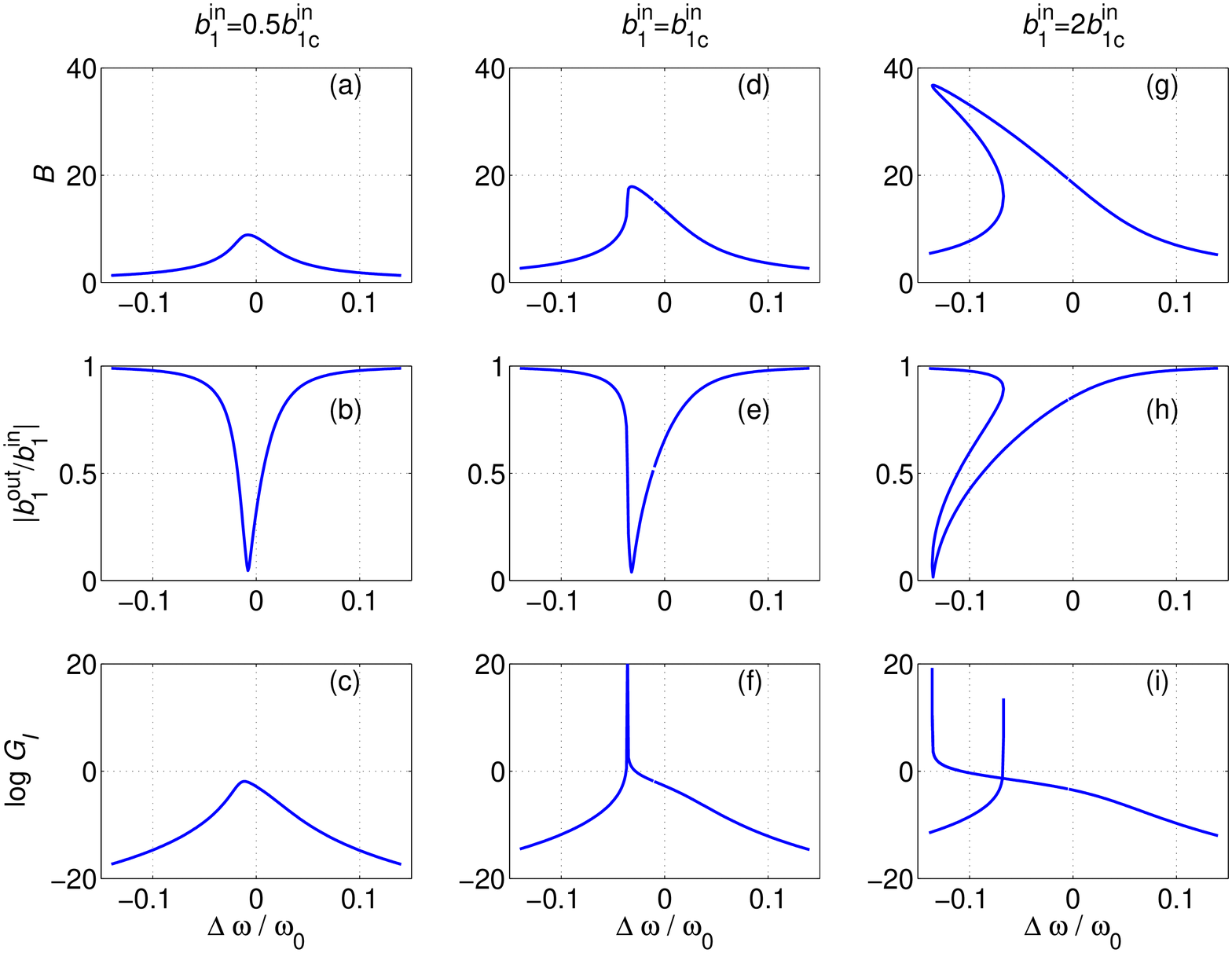}%
\caption{The cavity-mode amplitude $\left\vert B\right\vert $, the reflection
amplitude $\left\vert b_{1}^{out}/b_{1}^{in}\right\vert $, and the
intermodulation gain $G_{I}$ for vanishing offset frequency $\omega=0$ shown
for sub-critical case $b_{1}^{in}=0.5b_{1C}^{in}$, critical case $b_{1}%
^{in}=b_{1C}^{in}$, and above-critical case $b_{1}^{in}=2b_{1C}^{in}$. In all
cases $K=-10^{-4}\omega_{0}$, $\gamma_{1}=0.01\omega_{0},$ $\gamma
_{2}=1.1\gamma_{1}$, and $\gamma_{3}=0.01K/\sqrt{3}$. For $b_{1}^{in}%
>b_{1C}^{in}$ the response becomes multi-value function of frequency in some
frequency range.}%
\end{center}
\end{figure}

\subsection{Special operating points}

As a function of the pump frequency $\omega_{p}$, $B$ will have the form of
the distorted Lorentzian curve (see Fig. 2 (a), (d), and (g)) exhibited by
Duffing oscillators \cite{nayfeh,landau,yurke95}. The maximum of the response
curve occurs when $\partial E/\partial\omega_{p}=0$. This condition yields
\begin{equation}
\omega_{0}-\omega_{p}+KE=0\ ,
\end{equation}
that is, the peak of the resonance curve is shifted by an amount $KB^{2}$. The
points of instability where the system will switch from one of the two
bistable states to the other are located where $\partial\omega_{p}/\partial
E=0$. This condition is satisfied when
\begin{equation}
(\gamma+2\gamma_{3}E)^{2}=(K^{2}+\gamma_{3}^{2})E^{2}-(\omega_{0}-\omega
_{p}+2KE)^{2}\ . \label{eq:6.1.11}%
\end{equation}
When, in addition, $\partial^{2}\omega_{p}/\partial E^{2}=0$, the two points
of instability coalesce into a single point. The condition $\partial^{2}%
\omega_{p}/\partial E^{2}=0$ is satisfied when
\begin{equation}
6(K^{2}+\gamma_{3}^{2})E+4[(\omega_{0}-\omega_{p})K+\gamma\gamma_{3}]=0\ .
\label{eq:6.1.10}%
\end{equation}
Large parametric gain is achieved at points where the slope of $E$ with
respect to $\omega_{p}$ becomes infinite, but in order to remain stable it is
desirable to operate the reflection parametric amplifier near the critical
point with parameters chosen so that the Duffing curve does not have a
bistable region. It is a straightforward exercise to show that in order for
the resonance curve to have a critical point at which both
Eqs.~(\ref{eq:6.1.11}) and (\ref{eq:6.1.10}) are satisfied, one must have
\begin{equation}
|K|>\sqrt{3}\gamma_{3}\ . \label{K>}%
\end{equation}
At the critical point one has
\begin{equation}
E_{c}=\frac{2\gamma}{\sqrt{3}(|K|-\sqrt{3}\gamma_{3})} \label{eq:6.1.17}%
\end{equation}
and
\begin{equation}
\omega_{0}-\omega_{p}=-\gamma\frac{K}{|K|}\left[  \frac{4\gamma_{3}%
|K|+\sqrt{3}(K^{2}+\gamma_{3}^{2})}{K^{2}-3\gamma_{3}^{2}}\right]  \ .
\label{eq:6.1.18}%
\end{equation}
The incoming pump amplitude required for operation at the critical point is
given by
\begin{equation}
(b_{1c}^{in})^{2}=\frac{4}{3\sqrt{3}}\frac{\gamma^{3}(K^{2}+\gamma_{3}^{2}%
)}{\gamma_{1}(|K|-\sqrt{3}\gamma_{3})^{3}}\ .
\end{equation}
Thus, the input power required for driving the system into the threshold of
bistability (critical point) is increased in the presence of two-photon
losses. Moreover, when $\gamma_{3}$ exceeds the value of $|K|/\sqrt{3}$ the
bistability regime becomes inaccessible (see Eq. (\ref{K>})).

When $\gamma_{3}=0$ these reduce to
\begin{align}
E_{c}  &  =\frac{2\sqrt{3}\gamma}{3|K|}\ ,\\
\omega_{0}-\omega_{p}  &  =-\sqrt{3}\gamma\frac{K}{|K|}\ ,\\
(b_{1c}^{in})^{2}  &  =\frac{4}{3\sqrt{3}}\frac{\gamma^{3}}{\gamma_{1}|K|}\ .
\end{align}
In panel (d) of Fig. 2 the amplitude of the cavity mode as a function of
frequency has been plotted for the case when the incoming pump amplitude is
that of the critical pump amplitude. One sees that the line shape of the
cavity mode is vertical at a point on the lower side of the resonance. The
line shape of the reflected power is shown in panel (e) of Fig. 2.

\section{Linearization}

A linearized analysis is now performed in which the incoming signal and the
noise from the losses are regarded as small compared to the pump. To that end
we write
\begin{equation}
a_{1}^{in}=b_{1}^{in}e^{-i(\omega_{p}t+\psi_{1})}+c_{1}^{in}e^{-i\omega_{p}%
t}\ ,
\end{equation}%
\begin{equation}
a_{2}^{in}=c_{2}^{in}e^{-i\omega_{p}t}\ ,
\end{equation}%
\begin{equation}
a_{3}^{in}=c_{3}^{in}e^{-i\omega_{p}t}\ ,
\end{equation}%
\begin{equation}
a_{1}^{out}=b_{1}^{out}e^{-i(\omega_{p}t+\psi_{1})}+c_{1}^{out}e^{-i\omega
_{p}t}\ ,
\end{equation}%
\begin{equation}
a_{2}^{out}=b_{2}^{out}e^{-i\omega_{p}t}+c_{2}^{out}e^{-i\omega_{p}t}\ ,
\end{equation}%
\begin{equation}
a_{3}^{out}=b_{3}^{out}e^{-i\omega_{p}t}+c_{3}^{out}e^{-i\omega_{p}t}\ ,
\end{equation}
and
\begin{equation}
A=Be^{-i(\omega_{p}t+\phi_{B})}+ae^{-i\omega_{p}t}%
\end{equation}
where $B$, $b_{1}^{out}$, $b_{2}^{out}$ and $b_{3}^{out}$ constitute the
solution for the response of the system to a classical pump in the absence of
signal and noise. The properties of this solution have already been discussed
in Section 4. The $c_{1}^{in}$, $c_{2}^{in}$, $c_{3}^{in}$, $c_{1}^{out}$,
$c_{2}^{out}$, $c_{3}^{out}$, and $a$ are regarded as small and will be kept
only up to linear order. Substituting these into the equations of motion
yields
\begin{align}
\frac{da}{dt}  &  =-[i(\omega_{0}-\omega_{p})+\gamma]a-2(iK+\gamma_{3}%
)B^{2}a\nonumber\\
&  -(iK+\gamma_{3})B^{2}e^{-i2\phi_{B}}a^{\dagger}-\ i\sqrt{2\gamma_{1}%
}e^{i\phi_{1}}c_{1}^{in}\nonumber\\
&  -i\sqrt{2\gamma_{2}}e^{i\phi_{2}}c_{2}^{in}-i2\sqrt{\gamma_{3}}%
Be^{i(\omega_{p}t+\phi_{B}+\phi_{3})}c_{3}^{in}\ ,
\end{align}%
\begin{equation}
c_{1}^{out}-c_{1}^{in}=-i\sqrt{2\gamma_{1}}e^{-i\phi_{1}}a\ ,
\end{equation}%
\begin{equation}
c_{2}^{out}-c_{2}^{in}=-i\sqrt{2\gamma_{2}}e^{-i\phi_{2}}a\ ,
\end{equation}%
\begin{equation}
c_{3}^{out}-c_{3}^{in}=-i2\sqrt{\gamma_{3}}Be^{-i(\omega_{p}t+\phi_{B}%
+\phi_{3})}a\ .
\end{equation}

\section{Solving the linearized equation}

Introducing
\begin{equation}
W=i(\omega_{0}-\omega_{p})+\gamma+2(iK+\gamma_{3})B^{2} \ ,
\end{equation}%
\begin{equation}
V=(iK+\gamma_{3})B^{2}e^{-2i\phi_{B}}\ ,
\end{equation}
and
\begin{align}
F  &  = -i\sqrt{2\gamma_{1}}e^{i\phi_{1}}c_{1}^{in}-i\sqrt{2\gamma_{2}%
}e^{i\phi_{2}}c_{2}^{in}\nonumber\\
&  \ \ \ - \ i2\sqrt{\gamma_{3}}Be^{i(\omega_{p}t+\phi_{B}+\phi_{3})}%
c_{3}^{in}\ ,
\end{align}
the linearized equation of motion can be written in the form
\begin{equation}
\frac{da}{dt}+Wa+Va^{\dagger}=F\ .
\end{equation}
From this last equation one obtains
\begin{equation}
\frac{d^{2}a}{dt^{2}}+2\Re(W)\frac{da}{dt}+(|W|^{2}-|V|^{2})a=\Gamma(t)
\label{eq:r1}%
\end{equation}
where
\begin{equation}
\Gamma(t)=\frac{dF}{dt}+W^{\ast}F-VF^{\dagger}(t)\ . \label{eq:r2}%
\end{equation}
Writing
\begin{equation}
a=e^{-\lambda t}\ ,
\end{equation}
the characteristic equation for the homogenous equation is given by
\begin{equation}
\lambda^{2}-2\Re(\omega)\lambda+|W|^{2}-|V|^{2}=0\ .
\end{equation}
This has the two roots
\begin{equation}
\lambda_{0}=\Re(W)-\sqrt{\Re^{2}(W)-|W|^{2}+|V|^{2})}\ ,
\end{equation}%
\begin{equation}
\lambda_{1}=\Re(W)+\sqrt{\Re^{2}(W)-|W|^{2}+|V|^{2})}\ ,
\end{equation}
or
\begin{equation}
\lambda_{0}=\gamma+2\gamma_{3}B^{2}-\sqrt{(K^{2}+\gamma_{3}^{2})B^{4}%
-(\omega_{0}-\omega_{p}+2KB^{2})^{2}}\ ,
\end{equation}%
\begin{equation}
\lambda_{1}=\gamma+2\gamma_{3}B^{2}+\sqrt{(K^{2}+\gamma_{3}^{2})B^{4}%
-(\omega_{0}-\omega_{p}+2KB^{2})^{2}}\ .
\end{equation}
The root $\lambda_{0}$ is zero when Eq.~(\ref{eq:6.1.11}) is satisfied, that
is, one has critical slowing down at the points where the slope of $E$ with
respect to $\omega_{p}$ is infinite.

Introducing the Fourier transforms
\begin{equation}
a(t)=\frac{1}{\sqrt{2\pi}}\int_{-\infty}^{\infty}d\omega a(\omega)e^{-i\omega
t}\ ,
\end{equation}%
\begin{equation}
c_{1}(t)=\frac{1}{\sqrt{2\pi}}\int_{-\infty}^{\infty}d\omega c_{1}%
(\omega)e^{-i\omega t}\ ,
\end{equation}%
\begin{equation}
c_{2}(t)=\frac{1}{\sqrt{2\pi}}\int_{-\infty}^{\infty}d\omega c_{2}%
(\omega)e^{-i\omega t}\ ,
\end{equation}%
\begin{equation}
c_{3}(t)=\frac{1}{\sqrt{2\pi}}\int_{-\infty}^{\infty}d\omega c_{3}%
(\omega)e^{-i\omega t}\ ,
\end{equation}%
\begin{equation}
\Gamma(t)=\frac{1}{\sqrt{2\pi}}\int_{-\infty}^{\infty}d\omega\Gamma
(\omega)e^{-i\omega t}\ ,
\end{equation}
Eq.~(\ref{eq:r1}) yields
\begin{equation}
a(\omega)=\frac{\Gamma(\omega)}{-\omega^{2}-2i\omega\Re(W)+(|W|^{2}-|V|^{2}%
)}\ .
\end{equation}
In terms of the roots of the characteristic equation this can be written as
\begin{equation}
a(\omega)=\frac{\Gamma(\omega)}{(-i\omega+\lambda_{0})(-i\omega+\lambda_{1})}
\label{eq:9.1.17}%
\end{equation}
where
\begin{align}
\Gamma(\omega)  &  =-i\sqrt{2\gamma_{1}}[(-i\omega+W^{\ast})e^{i\phi_{1}}%
c_{1}^{in}(\omega)-Ve^{-i\phi_{1}}c_{1}^{in\dagger}(-\omega)]\nonumber\\
&  -\ i\sqrt{2\gamma_{2}}[(-i\omega+W^{\ast})e^{i\phi_{2}}c_{2}^{in}%
(\omega)-Ve^{-i\phi_{2}}c_{2}^{in\dagger}(-\omega)]\nonumber\\
&  -\ i2\sqrt{\gamma_{3}}B[(-i\omega+W^{\ast})e^{i(\phi_{B}+\phi_{3})}%
c_{3}^{in}(\omega_{p}+\omega)\nonumber\\
&  \ -\ Ve^{-i(\phi_{B}+\phi_{3})}c_{3}^{in\dagger}(\omega_{p}-\omega)]\ .
\end{align}

\subsection{The output field}

From Eq.~(\ref{eq:m2}) one obtains
\begin{equation}
c_{1}^{out}(\omega)=c_{1}^{in}(\omega)-i\sqrt{2\gamma_{1}}e^{-i\phi_{1}%
}a(\omega)\ .
\end{equation}
Substituting Eq.~(\ref{eq:9.1.17}) into this equation yields
\begin{align}
&  c_{1}^{out}(\omega)\nonumber\\
&  =\frac{(-i\omega+\lambda_{0})(-i\omega+\lambda_{1})c_{1}^{in}%
(\omega)-i\sqrt{2\gamma_{1}}e^{-i\phi_{1}}\Gamma(\omega)}{(-i\omega
+\lambda_{0})(-i\omega+\lambda_{1})}%
\end{align}
or
\begin{align}
&  c_{1}^{out}(\omega)\nonumber\\
&  =\frac{(-i\omega+\lambda_{0})(-i\omega+\lambda_{1})-2\gamma_{1}%
(-i\omega+W^{\ast})}{(-i\omega+\lambda_{0})(-i\omega+\lambda_{1})}c_{1}%
^{in}(\omega)\nonumber\\
&  +\ \frac{2\gamma_{1}Ve^{-i2\phi_{1}}}{(-i\omega+\lambda_{0})(-i\omega
+\lambda_{1})}c_{1}^{in\dagger}(-\omega)\nonumber\\
&  -\ \frac{2\sqrt{\gamma_{1}\gamma_{2}}(-i\omega+W^{\ast})e^{-i(\phi_{1}%
-\phi_{2})}}{(-i\omega+\lambda_{0})(-i\omega+\lambda_{1})}c_{2}^{in}%
(\omega)\nonumber\\
&  +\ \frac{2\sqrt{\gamma_{1}\gamma_{2}}Ve^{-i(\phi_{1}+\phi_{2})}}%
{(-i\omega+\lambda_{0})(-i\omega+\lambda_{1})}c_{2}^{in\dagger}(-\omega
)\nonumber\\
&  -\ \frac{2\sqrt{2\gamma_{1}\gamma_{3}}B(-i\omega+W^{\ast})e^{-i(\phi
_{1}-\phi_{B}-\phi_{3})}}{(-i\omega+\lambda_{0})(-i\omega+\lambda_{1})}%
c_{3}^{in}(\omega_{p}+\omega)\nonumber\\
&  +\frac{2\sqrt{2\gamma_{1}\gamma_{3}}BVe^{-i(\phi_{1}+\phi_{3}+\phi_{B})}%
}{(-i\omega+\lambda_{0})(-i\omega+\lambda_{1})}c_{3}^{in\dagger}(\omega
_{p}-\omega)\ . \label{eq:w1}%
\end{align}
The linearized solution to the equations of motion has now been obtained. We
will now evaluate the properties of this solution for various kinds of inputs.

\section{Parametric and intermodulation gain}

The parametric gain and the intermodulation gain are calculated by taking
$c_{1}^{in}(\omega)$ to represent a classical signal at frequency $\omega
_{p}+\omega$. Setting all other signal and noise inputs to zero,
Eq.~(\ref{eq:w1}) yields the following power gain for the reflected signal:
\begin{align}
G_{S}  &  \equiv\frac{|c_{1}^{out}(\omega)|^{2}}{|c_{1}^{in}(\omega)|^{2}%
}\nonumber\\
&  =\frac{|(-i\omega+\lambda_{0})(-i\omega+\lambda_{1})-2\gamma_{1}%
(-i\omega+W^{\ast})|^{2}}{(\omega^{2}+\lambda_{0}^{2})(\omega^{2}+\lambda
_{1}^{2})}\ .
\end{align}
When this quantity becomes greater than unity one has parametric amplification
of the signal.

As seen from Eq.~(\ref{eq:w1}), a signal $c_{in}(-\omega)$ injected at
frequency $\omega_{p}-\omega$ will generate an output signal at frequency
$\omega_{p}+\omega$. This frequency conversion is quantified by the
intermodulation conversion gain defined by
\begin{align}
G_{I}  &  \equiv\frac{|c_{1}^{out}(\omega)|^{2}}{|c_{1}^{in}(-\omega)|^{2}%
}\nonumber\\
&  =\frac{4\gamma_{1}^{2}|V|^{2}}{(\omega^{2}+\lambda_{0}^{2})(\omega
^{2}+\lambda_{1}^{2})}\ .
\end{align}
Since the output signal at $\omega_{p}+\omega$ is separated in frequency from
the input signal, the measurement of the intermodulation gain is a
particularly sensitive method for measuring the strength of the
nonlinearities. We note that, even without power gain, devices capable of
producing intermodulation signals are useful as mixers. When $\omega=0$, both
the expression for $G_{S}$ and the expression for $G_{I}$ will have
$\lambda_{0}^{2}\lambda_{1}^{2}$ in the denominator. As one approaches an
operating point where the slope of $E$ with respect to $\omega_{p}$ becomes
infinite, both the parametric gain and the intermodulation conversion gain
will diverge. Hence, it is near the instability points where the device can
exhibit large gains. Panels (c), (f), and (i) of Fig. 2 show the behavior of
the intermodulation gain as the pump amplitude is increased from half critical
(c) to critical (f) to twice critical (i) as a function of frequency. As
depicted in panel (f) at the critical point the intermodulation gain diverges.
Above critical, as shown in panel (i), the intermodulation gain diverges as
one approaches the points of infinite slope on the resonance curve (g).

\section{Noise squeezing}

Because of intermodulation gain, a parametric amplifier can establish
correlations \cite{yurke85} between the output at $\omega_{p}+\omega$ and
$\omega_{p}-\omega$. When delivered to a mixer whose local oscillator is
phase-locked to the pump these correlations can result in noise fluctuations
reduced below that which the mixer would see if the signal delivered to the
parametric amplifier were, instead, directly delivered to the mixer. This
noise reduction is called squeezing, and it can occur with either thermal or
quantum noise \cite{movshovich90}. We now obtain expressions that will allow
one to calculate the degree of thermal or quantum noise squeezing. For such a
calculation the $c_{1}^{in}$, $c_{2}^{in}$, $c_{3}^{in}$, $c_{1}^{out}$,
$c_{2}^{out}$, $c_{3}^{out}$ are again treated as quantum mechanical operators
satisfying commutation relations of the form Eqs.~(\ref{eq:c1}) and
(\ref{eq:c2}).

The output of a mixer, operated in the homodyne mode in which the local
oscillator frequency $\omega_{LO}$ and the pump frequency $\omega_{p}$ are
equal and in which the input at the signal frequency $\omega_{LO}+\omega$ and
at the image frequency $\omega_{LO}-\omega$ are both regarded as signal, is
given by \cite{yurke89}
\begin{equation}
I_{D}(\omega)=c_{1}^{out\dagger}(-\omega)e^{-i\phi_{LO}}+c_{1}^{out}%
(\omega)e^{i\phi_{LO}} \label{eq:12.1}%
\end{equation}
where $\phi_{LO}$ is the local oscillator phase. To evaluate the mean value
and the power spectrum for the homodyne detector output it is necessary to
specify the density matrix for the signal and noise entering the parametric
amplifier. Here we consider the case when these inputs consist of Nyquist
noise. In this case one has
\begin{equation}
\langle c_{i}^{in}(\omega)\rangle=0\ ,
\end{equation}%
\begin{equation}
\langle c_{i}^{in\dagger}(\omega)c_{j}^{in}(\omega^{\prime})\rangle
=\frac{e^{-\beta_{i}\hbar\omega_{p}}}{1-e^{-\beta_{i}\hbar\omega_{p}}}%
\delta_{i,j}\delta(\omega-\omega^{\prime})\ , \label{eq:11.2.2}%
\end{equation}
and
\begin{equation}
\langle c_{i}^{in}(\omega)c_{j}^{in}(\omega^{\prime})\rangle=0\ .
\end{equation}
Here
\begin{equation}
\beta_{i}=\frac{1}{k_{B}T_{i}}\ ,
\end{equation}
where $k_{B}$ is Boltzmann's constant and $T_{i}$ is the absolute temperature
of the bath for which $c_{i}^{in}(\omega)$ is the incoming mode. We thus allow
each of the baths to be at a different temperature. In writing
Eq.~(\ref{eq:11.2.2}) we have made the approximation that the frequencies
$\omega$ of interest are small compared to $\omega_{p}$.

Because Eq.~(\ref{eq:w1}) is linear in the $c_{i}^{in}(\omega)$ it is evident
that
\begin{equation}
\langle I_{D}(\omega)\rangle=0\ ,
\end{equation}
that is, the homodyne detector output consists of noise fluctuations with zero
mean. Because of the boson commutation relations, one has
\begin{equation}
\langle I_{D}^{\dagger}(\omega)I_{D}(\omega^{\prime})\rangle=P(\omega
)\delta(\omega-\omega^{\prime}) \label{eq:p1}%
\end{equation}
where $P(\omega)$ is the noise power spectrum of the homodyne detector output.
Equation~(\ref{eq:w1}) can be rewritten as
\begin{align}
c_{1}^{out}(\omega)  &  =A_{1}(\omega)c_{1}^{in}(\omega)+B_{1}(\omega
)c_{1}^{in\dagger}(-\omega)\nonumber\\
&  +\ A_{2}(\omega)c_{2}^{in}(\omega)+B_{2}(\omega)c_{2}^{in\dagger}%
(-\omega)\nonumber\\
&  +\ A_{3}(\omega)c_{3}^{in}(\omega_{p}+\omega)+B_{3}(\omega)c_{3}%
^{in\dagger}(\omega_{p}-\omega) \label{eq:12.3}%
\end{align}
where
\begin{equation}
A_{1}(\omega)=\frac{(-i\omega+\lambda_{0})(-i\omega+\lambda_{1})-2\gamma
_{1}(-i\omega+W^{\ast})}{(-i\omega+\lambda_{0})(-i\omega+\lambda_{1})}\ ,
\end{equation}%
\begin{equation}
B_{1}(\omega)=\frac{2\gamma_{1}Ve^{-i2\phi_{1}}}{(-i\omega+\lambda
_{0})(-i\omega+\lambda_{1})}\ ,
\end{equation}%
\begin{equation}
A_{2}(\omega)=-\frac{2\sqrt{\gamma_{1}\gamma_{2}}(-i\omega+W^{\ast}%
)e^{-i(\phi_{1}-\phi_{2})}}{(-i\omega+\lambda_{0})(-i\omega+\lambda_{1})}\ ,
\end{equation}%
\begin{equation}
B_{2}(\omega)=\frac{2\sqrt{\gamma_{1}\gamma_{2}}Ve^{-i(\phi_{1}+\phi_{2})}%
}{(-i\omega+\lambda_{0})(-i\omega+\lambda_{1})}\ ,
\end{equation}%
\begin{equation}
A_{3}(\omega)=-\ \frac{2\sqrt{2\gamma_{1}\gamma_{3}}B(-i\omega+W^{\ast
})e^{-i(\phi_{1}-\phi_{B}-\phi_{3})}}{(-i\omega+\lambda_{0})(-i\omega
+\lambda_{1})}\ ,
\end{equation}%
\begin{equation}
B_{3}(\omega)=\frac{2\sqrt{2\gamma_{1}\gamma_{3}}BVe^{-i(\phi_{1}+\phi
_{3}+\phi_{B})}}{(-i\omega+\lambda_{0})(-i\omega+\lambda_{1})}\ .
\end{equation}
Substituting Eq.~(\ref{eq:12.3}) into Eq.~(\ref{eq:12.1}), evaluating $\langle
I_{D}^{\dagger}(\omega)I_{D}(\omega^{\prime})\rangle$ using
Eq.~(\ref{eq:11.2.2}), and then reading off the power spectrum using
Eq.~(\ref{eq:p1}), one obtains
\begin{align}
P(\omega)  &  =|e^{-i\phi_{LO}}A_{1}^{\ast}(\omega)+e^{i\phi_{LO}}%
B_{1}(-\omega)|^{2}\frac{e^{-\beta_{1}\hbar\omega_{p}}}{1-e^{-\beta_{1}%
\hbar\omega_{p}}}\nonumber\\
&  +\ |e^{i\phi_{LO}}A_{1}(-\omega)+e^{-i\phi_{LO}}B_{1}^{\ast}(\omega
)|^{2}\frac{1}{1-e^{-\beta_{1}\hbar\omega_{p}}}\nonumber\\
&  +\ |e^{-i\phi_{LO}}A_{2}^{\ast}(\omega)+e^{i\phi_{LO}}B_{2}(-\omega
)|^{2}\frac{e^{-\beta_{2}\hbar\omega_{p}}}{1-e^{-\beta_{2}\hbar\omega_{p}}%
}\nonumber\\
&  +\ |e^{i\phi_{LO}}A_{2}(-\omega)+e^{-i\phi_{LO}}B_{2}^{\ast}(\omega
)|^{2}\frac{1}{1-e^{-\beta_{2}\hbar\omega_{p}}}\nonumber\\
&  +\ |e^{-i\phi_{LO}}A_{3}^{\ast}(\omega)+e^{i\phi_{LO}}B_{3}(-\omega
)|^{2}\frac{e^{-\beta_{3}\hbar\omega_{p}}}{1-e^{-\beta_{3}\hbar\omega_{p}}%
}\nonumber\\
&  +\ |e^{i\phi_{LO}}A_{3}(-\omega)+e^{-i\phi_{LO}}B_{3}^{\ast}(\omega
)|^{2}\frac{1}{1-e^{-\beta_{3}\hbar\omega_{p}}} \ . \label{P(omega)}%
\end{align}
This formula may be used to compute the noise-power spectrum for any local
oscillator phase $\phi_{LO}$ and any set of device parameters. It is useful to
consider the case when there is no incoming pump field and the input field and
loss baths are all at zero temperature. In this case the field reflected off
the input port of the amplifier will consist of vacuum fluctuations, that is,
\begin{equation}
c_{1}^{out}(\omega)|0\rangle=0\ ,
\end{equation}
and one obtains
\begin{equation}
\langle I_{D}^{\dagger}(\omega)I_{D}(\omega^{\prime})\rangle=\delta
(\omega-\omega^{\prime})
\end{equation}
or
\begin{equation}
P(\omega)=1\ .
\end{equation}
This sets the vacuum noise level for the conventions we are using. As is shown
in Fig. 3 and as will be illustrated with specific examples in the next
section, under suitable circumstances it is possible to obtain reflected
signals whose noise-power spectrum $P(\omega)$, for certain local oscillator
phase settings, is less than 1. Such signals are said to be squeezed below the
vacuum noise level. Figure 3 shows the minimum value of $P(\omega)$ as a
function of the amplitude of the incoming pump $b_{1}^{in}$ when the bath
temperatures are all zero, the strength of the Kerr nonlinearity is chosen to
be $K=5\omega_{0}$, and the coupling strength of the signal to the cavity mode
is taken to be $\gamma_{1}=0.0001\omega_{0}$. The pump frequency has been
chosen to be that of the critical pump frequency. The solid line is the
lossless case when $\gamma_{2}=\gamma_{3}=0$. In this case complete squeezing
$P(0)=0$ is possible when the incoming pump power is at the critical value,
that is, $|b_{1}^{in}/b_{1c}^{in}|=1$. The dashed curve represents the case
when the linear dissipation is given by $\gamma_{2}=5\gamma_{1}$ and the
nonlinear dissipation $\gamma_{3}$ is zero. In general, the presence of linear
dissipation reduces the amount of achievable squeezing. The dotted line
depicts the case when the linear dissipation is zero and the nonlinear
dissipation is given by $\gamma_{3}=0.5K/\sqrt{3}$. Nonlinear dissipation also
tends to reduce the amount of squeezing that can be produced. However,
nonlinear dissipation can produce some squeezing in the absence of a Kerr
medium, as we discuss below.%

\begin{figure}
[ptb]
\begin{center}
\includegraphics[
height=2.6697in,
width=3.2396in
]%
{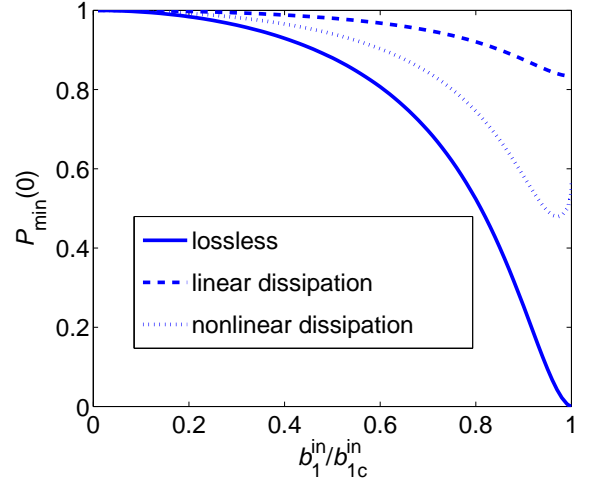}%
\caption{Examples of achievable degree of squeezing, $P_{min}(0)$ vs.
$b_{1}^{in}/b_{1c}^{in}$.\ In all plots $T_{1}=T_{2}=T_{3}=0$, $K=5\omega_{0}%
$, $\gamma_{1}=0.0001\omega_{0}$, and the pump frequency $\omega_{p}$ obtains
its critical value, as given by Eq. \ref{eq:6.1.18}. The solid line represents
the lossless case, where $\gamma_{2}=\gamma_{3}=0$. \ The dashed line
represents the case of linear dissipation where $\gamma_{2}=5\gamma_{1}$ and
$\gamma_{3}=0$, while the dotted line represents the case of nonlinear
dissipation where $\gamma_{2}=0$ and $\gamma_{3}=0.5K/\sqrt{3}$.}%
\end{center}
\end{figure}

\subsection{Special cases}

It is instructive to evaluate Eq. (\ref{P(omega)}) for a few specific cases.
In particular, the power spectrum at $\omega=0$ reduces to
\begin{align}
P(0)  &  =|e^{-i\phi_{LO}}A_{1}^{\ast}(0)+e^{i\phi_{LO}}B_{1}(0)|^{2}%
\coth\left(  \frac{\hbar\omega_{p}}{2k_{B}T_{1}}\right)
\nonumber\label{eq:12.1.1}\\
&  +\ |e^{-i\phi_{LO}}A_{2}^{\ast}(0)+e^{i\phi_{LO}}B_{2}(0)|^{2}\coth\left(
\frac{\hbar\omega_{p}}{2k_{B}T_{2}}\right) \nonumber\\
&  +\ |e^{-i\phi_{LO}}A_{3}^{\ast}(0)+e^{i\phi_{LO}}B_{3}(0)|^{2}\coth\left(
\frac{\hbar\omega_{p}}{2k_{B}T_{3}}\right)  \ .
\end{align}
Further simplification results from considering the case when the internal
losses are all zero. In this case $\gamma_{2}=0$ and $\gamma_{3}=0$. Thus, the
terms in Eq.~(\ref{eq:12.1.1}) corresponding to internally generated noise
vanish and one obtains
\begin{equation}
P(0)=|e^{-i\phi_{LO}}A_{1}^{\ast}(0)+e^{i\phi_{LO}}B_{1}(0)|^{2}\coth\left(
\frac{\hbar\omega_{p}}{2k_{B}T_{1}}\right)  \ .
\end{equation}
Writing
\begin{align}
A_{1}(0)  &  =|A_{1}(0)|e^{i\phi_{A}}\ ,\\
B_{1}(0)  &  =|B_{1}(0)|e^{i\phi_{B}}\ ,
\end{align}
one has
\begin{equation}
P(0)=\left\vert |A_{1}(0)|+|B_{1}(0)|e^{i\psi}\right\vert ^{2}\coth\left(
\frac{\hbar\omega_{p}}{2k_{B}T_{1}}\right)
\end{equation}
where
\begin{equation}
\psi=2\phi_{LO}-\phi_{A}+\phi_{B}\ . \label{eq:l1}%
\end{equation}
$P(0)$ is maximized when the local oscillator phase $\phi_{LO}$ is chosen so
that
\begin{equation}
e^{i\psi}=1\ . \label{eq:l2}%
\end{equation}
In this case one obtains
\begin{equation}
P_{max}(0)=(|A_{1}(0)|+|B_{1}(0)|)^{2}\coth\left(  \frac{\hbar\omega_{p}%
}{2k_{B}T_{1}}\right)  \ . \label{eq:12.2.7}%
\end{equation}
$P(0)$ is minimized when the local oscillator phase $\phi_{LO}$ is chosen so
that
\begin{equation}
e^{i\psi}=-1\ . \label{eq:l3}%
\end{equation}
In this case one obtains
\begin{equation}
P_{min}(0)=(|A_{1}(0)|-|B_{1}(0)|)^{2}\coth\left(  \frac{\hbar\omega_{p}%
}{2k_{B}T_{1}}\right)  \ . \label{eq:12.2.9}%
\end{equation}
From Eqs.~(\ref{eq:l1}), (\ref{eq:l2}), and (\ref{eq:l3}) it follows that the
local oscillator phase at which the spectral density $P(0)$ is minimized
differs by $\pi/2$ from the local oscillator phase at which the spectral
density is maximized, that is, the signal components minimizing and maximizing
the spectral density are in quadrature. It is straightforward to show that
\begin{equation}
|A_{1}(0)|^{2}-|B_{1}(0)|^{2}=1\ .
\end{equation}
From this equation and Eqs.~(\ref{eq:12.2.7}) and (\ref{eq:12.2.9}), one
obtains
\begin{equation}
P_{max}(0)P_{min}(0)=\coth^{2}\left(  \frac{\hbar\omega_{p}}{2k_{B}T_{1}%
}\right)  \ .
\end{equation}
Hence, in the case of no loss, the degree of amplification and the degree of
deamplification of the noise are the same. For this case one also has
\begin{equation}
|B_{1}(0)|=\frac{2\gamma_{1}KB^{2}}{(\omega_{0}-\omega_{p}+2KB^{2})^{2}%
+\gamma_{1}^{2}-K^{2}B^{4}}\ .
\end{equation}
Note that $|B_{1}(0)|$ can be made arbitrarily large by choosing the pump
frequency $\omega_{p}$ and the pump amplitude $B$ such that the denominator
goes to zero. From this it follows that $P_{max}$ can be made arbitrarily
large and $P_{min}$ can be made arbitrarily small. In practice, higher order
terms responsible for pump depletion will limit how big $P_{max}$ can be made.

Noise from the losses and gain saturation will limit the maximum degree of
noise squeezing to be below that calculated here. This is demonstrated in
Figure 3, where the achievable degree of squeezing $P_{min}(0)$ is plotted as
a function of $b_{1}^{in}/b_{1c}^{in}$ for some examples in which the linear
or nonlinear loss is taken to be nonzero.

The presence of a two-photon loss allows for squeezing even in the absence of
a Kerr nonlinearity \cite{ezaki99}. Setting $K = 0$, $\gamma_{2} = 0$, and
$\omega_{0} = \omega_{p}$, the greatest degree of squeezing of the output
field of the cavity occurs when $\gamma_{1} = 3\gamma_{3}B^{2}$. At this
operating point $P(0) = 2/3$ and the local oscillator phase must be adjusted
so that $\cos(2\phi_{LO} - 2\phi_{B} - 2\phi_{1}) = 1$.

\section{Conclusions}

We have presented an analysis of a cavity parametric amplifier employing a
Kerr nonlinearity but which also possesses a two-photon loss. We have obtained
expressions for the pump amplitude inside the cavity and the reflected pump
amplitude for the case when pump saturation can be neglected. We have obtained
expressions for the classical gain and the intermodulation gain. These
expressions are useful for determining model parameters from experimental
data. We have found that in the presence of two photon losses the injected
power required for driving the system into the bistable regime is increased.
Moreover, this regime becomes inaccessible when $\gamma_{3}$ exceeds the value
of $|K|/\sqrt{3}$.

We have also obtained expressions from which one can compute the degree of
squeezing that the device exhibits. Both the linear and nonlinear loss tend to
degrade the amount of squeezing that can be achieved, although even without a
Kerr nonlinearity a modest amount of squeezing can be achieved by the
two-photon loss.

\appendix

\section{Nonlinear Kinetic Inductance in Transmission Line Resonator}

Consider a lossless linear transmission line with length $l$ extending along
the $x$-axis. \ Let $q\left(  x,t\right)  $ be the charge density per unit
length and define \cite{yurke84}%

\begin{equation}
Q\left(  x,t\right)  =\int\limits_{x}^{\infty}dx^{\prime}q\left(  x^{\prime
},t\right)  .
\end{equation}

Thus $q=-\partial Q/\partial x$, and the voltage across the transmission line
is given by%

\begin{equation}
V\left(  x,t\right)  =-\frac{1}{C}\frac{\partial Q}{\partial x},
\end{equation}

where $C$ is the capacitance per unit length along the transmission line,
whereas the current is given by%

\begin{equation}
I\left(  x,t\right)  =\frac{\partial Q}{\partial t}.
\end{equation}

The Lagrangian $\mathcal{L}$ of the system reads%

\begin{align}
\mathcal{L}  &  =\frac{1}{2}\int\limits_{0}^{l}dx\left[  LI^{2}-CV^{2}\right]
\\
&  =\frac{1}{2}\int\limits_{0}^{l}dx\left[  L\left(  \frac{\partial
Q}{\partial t}\right)  ^{2}-\frac{1}{C}\left(  \frac{\partial Q}{\partial
x}\right)  ^{2}\right]  ,\nonumber
\end{align}

where $L$ is the inductance per unit length along the transmission line. \ The
open ends at $x=0$ and $x=l$ impose boundary conditions of vanishing current.

We assume the case of a nonuniform transmission line, where both $C$ and $L$
may depend on $x$. Moreover, the inductance $L$ depends on the current $I$
according to Eq. (\ref{L}) as a result of nonlinear kinetic inductance.

As a basis for expanding $Q\left(  x,t\right)  $ as%

\begin{equation}
Q\left(  x,t\right)  =\sum_{n}q_{n}\left(  t\right)  u_{n}\left(  x\right)  ,
\label{Q(x,t)}%
\end{equation}
we use the solutions of%

\begin{equation}
\frac{d}{dx}\left(  \frac{1}{C}\frac{du_{n}}{dx}\right)  =-\omega_{n}%
^{2}Lu_{n}, \label{we u}%
\end{equation}
with the boundary conditions of vanishing current%

\begin{equation}
u_{n}\left(  0\right)  =u_{n}\left(  l\right)  =0.
\end{equation}
We assume that the functions $u_{n}\left(  x\right)  $ are chosen to be real.

For the case of a uniform transmission line resonator, where both $C$ and $L$
are independent on $x$, one has $\omega_{n}=n\pi/l\sqrt{LC}$, where $n$ is an
integer. In the nonlinear regime, however, such an equally spaced spectrum may
lead to strong inter-mode coupling, where harmonics and sub-harmonics of a
driven mode excite other modes. In the present work we assume that such
inter-modes effects are avoided by employing a nonuniform resonator. \ This
allows us to consider only the mode in the resonator which is driven externally.

Using expansion (\ref{Q(x,t)})%

\begin{align}
\mathcal{L}  &  =\frac{1}{2}\sum_{n}\sum_{m}\dot{q}_{n}\dot{q}_{m}%
\int\limits_{0}^{l}dxL_{0}u_{n}u_{m}\\
&  -\frac{1}{2}\sum_{n}\sum_{m}q_{n}q_{m}\int\limits_{0}^{l}dx\frac{1}{C}%
\frac{du_{n}}{dx}\frac{du_{m}}{dx}+\Delta\mathcal{L},\nonumber
\end{align}
where%

\begin{align}
\Delta\mathcal{L}  &  =\frac{1}{2I_{c}^{2}}\sum_{n^{\prime},n^{\prime\prime
},n^{\prime\prime\prime},n^{\prime\prime\prime\prime}}\dot{q}_{n^{\prime}}%
\dot{q}_{n^{\prime\prime}}\dot{q}_{n^{\prime\prime\prime}}\dot{q}%
_{n^{\prime\prime\prime\prime}}\nonumber\\
&  \times\int\limits_{0}^{l}dx\ u_{n^{\prime}}u_{n^{\prime\prime}}%
u_{n^{\prime\prime\prime}}u_{n^{\prime\prime\prime\prime}}\Delta L.
\end{align}

We first treat the linear part. \ Consider equation (\ref{we u}) for $u_{n}$
multiplied by $u_{m}$ and equation (\ref{we u}) for $u_{m}$ multiplied by
$u_{n}$%

\begin{equation}
u_{m}\frac{d}{dx}\left(  \frac{1}{C}\frac{du_{n}}{dx}\right)  =-\omega_{n}%
^{2}L_{0}u_{n}u_{m},
\end{equation}

\begin{equation}
u_{n}\frac{d}{dx}\left(  \frac{1}{C}\frac{du_{m}}{dx}\right)  =-\omega_{m}%
^{2}L_{0}u_{n}u_{m}.
\end{equation} Subtracting%

\begin{equation}
\frac{d}{dx}\left(  u_{m}\frac{1}{C}\frac{du_{n}}{dx}-u_{n}\frac{1}{C}%
\frac{du_{m}}{dx}\right)  =\left(  \omega_{m}^{2}-\omega_{n}^{2}\right)
L_{0}u_{n}u_{m},
\end{equation}
and integrating from $x=0$ to $x=l$ one obtains%

\begin{equation}
\left(  \omega_{m}^{2}-\omega_{n}^{2}\right)  \int\limits_{0}^{l}dxL_{0}%
u_{n}u_{m}=0.
\end{equation}

In general, it can be easily shown that the spectrum of a finite
one-dimensional resonator having vanishing current boundary conditions is
non-degenerate. Thus, by requiring that the functions $u_{n}\left(  x\right)
$ are normalized one obtains%

\begin{equation}
\int\limits_{0}^{l}dxL_{0}u_{n}u_{m}=\delta_{nm}.
\end{equation}
Moreover, integrating by parts and using (\ref{we u}) and the boundary
conditions one obtains%

\begin{equation}
\int\limits_{0}^{l}dx\frac{1}{C}\frac{du_{n}}{dx}\frac{du_{m}}{dx}=\omega
_{n}^{2}\int\limits_{0}^{l}dxL_{0}u_{n}u_{m}=\omega_{n}^{2}\delta_{nm}.
\end{equation}

Thus%

\begin{equation}
\mathcal{L}=\frac{1}{2}\sum_{n}\left(  \dot{q}_{n}^{2}-\omega_{n}^{2}q_{n}%
^{2}\right)  +\Delta\mathcal{L}.\nonumber
\end{equation}

The Euler-Lagrange equation is given by%

\begin{equation}
\frac{d}{dt}\left(  \frac{\partial\mathcal{L}}{\partial\dot{q}_{n}}\right)
-\frac{\partial\mathcal{L}}{\partial q_{n}}=0, \label{Euler Lagrange}%
\end{equation}
thus%

\begin{equation}
\ddot{q}_{n}+\omega_{n}^{2}q_{n}+\frac{d}{dt}\left(  \frac{\partial
\Delta\mathcal{L}}{\partial\dot{q}_{n}}\right)  =0.
\end{equation}

The variable canonically conjugate to $q_{n}$ is%

\begin{equation}
p_{n}=\frac{\partial\mathcal{L}}{\partial\dot{q}_{n}}=\dot{q}_{n}%
+\frac{\partial\Delta\mathcal{L}}{\partial\dot{q}_{n}} \label{can conj}%
\end{equation}

The Hamiltonian is given by%

\begin{align}
\mathcal{H}  &  =\sum_{n}p_{n}\dot{q}_{n}-\mathcal{L}\\
&  =\frac{1}{2}\sum_{n}\left[  p_{n}^{2}-\left(  \frac{\partial\Delta
\mathcal{L}}{\partial\dot{q}_{n}}\right)  ^{2}+\omega_{n}^{2}q_{n}^{2}\right]
-\Delta\mathcal{L}.\nonumber
\end{align}

To first order in $\Delta\mathcal{L}$ (or in $\Delta L$)%

\begin{equation}
\mathcal{H}=\mathcal{H}_{0}+\mathcal{V},
\end{equation}
where%

\begin{equation}
\mathcal{H}_{0}=\frac{1}{2}\sum_{n}\left(  p_{n}^{2}+\omega_{n}^{2}q_{n}%
^{2}\right)  , \label{H_0}%
\end{equation}
and%

\begin{align}
\mathcal{V}  &  =-\frac{1}{2I_{c}^{2}}\sum_{n^{\prime},n^{\prime\prime
},n^{\prime\prime\prime},n^{\prime\prime\prime\prime}}p_{n^{\prime}%
}p_{n^{\prime\prime}}p_{n^{\prime\prime\prime}}p_{n^{\prime\prime\prime\prime
}}\\
&  \times\int\limits_{0}^{l}dx\ u_{n^{\prime}}u_{n^{\prime\prime}}%
u_{n^{\prime\prime\prime}}u_{n^{\prime\prime\prime\prime}}\Delta L.\nonumber
\end{align}

To quantize the problem the variables $q_{n}$ and $p_{n}$ are regarded as
operators satisfying the following commutation relations%

\begin{equation}
\left[  q_{n},p_{m}\right]  \equiv q_{n}p_{m}-p_{m}q_{n}=i\hbar\delta_{n,m}.
\label{q,p}%
\end{equation}

\begin{equation}
\left[  q_{n},q_{m}\right]  =\left[  p_{n},p_{m}\right]  =0.
\label{(q,q)(p,p)}%
\end{equation}

The Boson annihilation and creation operators are defined as%

\begin{equation}
A_{n}=\frac{e^{i\omega_{n}t}}{\sqrt{2\hbar}}\left(  \sqrt{\omega_{n}}%
q_{n}+\frac{i}{\sqrt{\omega_{n}}}p_{n}\right)  , \label{a}%
\end{equation}

\begin{equation}
A_{n}^{\dag}=\frac{e^{-i\omega_{n}t}}{\sqrt{2\hbar}}\left(  \sqrt{\omega_{n}%
}q_{n}-\frac{i}{\sqrt{\omega_{n}}}p_{n}\right)  , \label{a+}%
\end{equation}

The inverse transformation is given by%

\begin{equation}
q_{n}=\sqrt{\frac{\hbar}{2\omega_{n}}}\left(  A_{n}^{\dag}e^{i\omega_{n}%
t}+A_{n}e^{-i\omega_{n}t}\right)  , \label{inv_q}%
\end{equation}

\begin{equation}
p_{n}=i\sqrt{\frac{\hbar\omega_{n}}{2}}\left(  A_{n}^{\dag}e^{i\omega_{n}%
t}-A_{n}e^{-i\omega_{n}t}\right)  . \label{inv_p}%
\end{equation}

The commutation relations for the operators $A_{n}$ and $A_{n}^{\dag}$ are
derived directly from (\ref{q,p}) and (\ref{(q,q)(p,p)})%

\begin{equation}
\left[  A_{n},A_{m}^{\dag}\right]  =\delta_{n,m}.
\end{equation}

\begin{equation}
\left[  A_{n},A_{m}\right]  =\left[  A_{n}^{\dag},A_{m}^{\dag}\right]  =0.
\label{[a,a],[a,a+]}%
\end{equation}

Using (\ref{inv_q}) and (\ref{inv_p}), the Hamiltonian (\ref{H_0}) can be
expressed as%

\begin{equation}
\mathcal{H}_{0}=\sum_{n}\hbar\omega_{n}\left(  A_{n}^{\dag}A_{n}+\frac{1}%
{2}\right)  .
\end{equation}

The current operator is given by%

\begin{equation}
I\left(  x,t\right)  =\frac{\partial Q}{\partial t}=i\sum_{n}\sqrt{\frac
{\hbar\omega_{n}}{2}}\left(  A_{n}^{\dag}e^{i\omega_{n}t}-A_{n}e^{-i\omega
_{n}t}\right)  u_{n}\left(  x\right)  . \label{I(x)}%
\end{equation}

The voltage operator is given by%

\begin{equation}
V\left(  x,t\right)  =-\frac{1}{C}\sum_{n}\sqrt{\frac{\hbar}{2\omega_{n}}%
}\left(  A_{n}^{\dag}e^{i\omega_{n}t}+A_{n}e^{-i\omega_{n}t}\right)
\frac{du_{n}}{dx}, \label{V(x)}%
\end{equation}

The terms $p_{n^{\prime}}p_{n^{\prime\prime}}p_{n^{\prime\prime\prime}%
}p_{n^{\prime\prime\prime\prime}}$ contain in general terms oscillating
rapidly at frequencies on the order of the frequencies in the resonator
spectrum. \ In the rotating wave approximation (RWA) these terms are neglected
since their effect on the dynamics on a time scale much longer compared to
typical oscillation period is negligibly small and only stationary terms
remain. Thus in the expression of $\mathcal{V}$ only terms of the type
$p_{n^{\prime}}^{2}p_{n^{\prime\prime}}^{2}$ contain stationary terms, which
are given by%

\begin{equation}
p_{n^{\prime}}^{2}p_{n^{\prime\prime}}^{2}\simeq\frac{\hbar\omega_{n^{\prime}%
}}{2}\frac{\hbar\omega_{n^{\prime\prime}}}{2}\left(  1+2A_{n^{\prime}}^{\dag
}A_{n^{\prime}}\right)  \left(  1+2A_{n^{\prime\prime}}^{\dag}A_{n^{\prime
\prime}}\right)  .
\end{equation}

The constant term can be disregarded since it only gives rise to a constant
phase factor. \ Moreover, the terms $A_{n^{\prime}}^{\dag}A_{n^{\prime}}$ and
$A_{n^{\prime\prime}}^{\dag}A_{n^{\prime\prime}}$ that give rise to frequency
shift can be absorbed into $\mathcal{H}_{0}$. \ Thus in the RWA the
perturbation $\mathcal{V}$ contain only terms of the type $A_{n^{\prime}%
}^{\dag}A_{n^{\prime}}A_{n^{\prime\prime}}^{\dag}A_{n^{\prime\prime}}$%

\begin{equation}
\mathcal{V}=\frac{\hbar}{2}\sum_{n^{\prime}}K_{n^{\prime}}\left(
A_{n^{\prime}}^{\dag}A_{n^{\prime}}\right)  ^{2}+\hbar\sum_{n^{\prime}\neq
n^{\prime\prime}}\lambda_{n^{\prime}n^{\prime\prime}}A_{n^{\prime}}^{\dag
}A_{n^{\prime}}A_{n^{\prime\prime}}^{\dag}A_{n^{\prime\prime}},
\end{equation}

where%

\begin{equation}
K_{n^{\prime}}=-\frac{\hbar\omega_{n^{\prime}}^{2}}{I_{c}^{2}}\int
\limits_{0}^{l}dx\ u_{n^{\prime}}^{4}\Delta L, \label{K_n'}%
\end{equation}

and%

\begin{equation}
\lambda_{n^{\prime}n^{\prime\prime}}=-\frac{3\hbar\omega_{n^{\prime}}%
\omega_{n^{\prime\prime}}}{I_{c}^{2}}\int\limits_{0}^{l}dx\ u_{n^{\prime}}%
^{2}u_{n^{\prime\prime}}^{2}\Delta L. \label{lambda_n'n''}%
\end{equation}

\section{Nonlinear losses associated with the kinetic inductance}

Here we derive expressions or the linear and nonlinear loss coefficients
$\gamma_{2}$ and $\gamma_{3}$ in terms of the parameters that characterize the
resistive loss associated with the kinetic inductance. This is accomplished by
obtaining an expression for the rate of energy loss in the cavity in terms of
the model parameters and comparing it with the expression for the power
dissipated due to the resistance associated with the kinetic inductance.

The equation of motion for the resonator Hamiltonian $H_{r}$ (Eq.
(\ref{H_r}))
\begin{equation}
i\hbar\frac{dH_{r}}{dt}=[H_{r},H] \label{eq:b2}%
\end{equation}
yields
\begin{align}
\frac{dH_{r}}{dt}  &  =-i\hbar\omega_{0}\int d\omega\lbrack\kappa
_{1}A^{\dagger}a_{1}(\omega)-\kappa_{1}^{\ast}a_{1}^{\dagger}(\omega
)A]\nonumber\\
&  -\ i\hbar\omega_{0}\int d\omega\lbrack\kappa_{2}A^{\dagger}a_{2}%
(\omega)-\kappa_{2}^{\ast}a_{2}^{\dagger}(\omega)A]\nonumber\\
&  -\ 2i\hbar\omega_{0}\int d\omega\lbrack\kappa_{3}A^{\dagger}A^{\dagger
}a_{3}(\omega)-\kappa_{3}^{\ast}a_{3}^{\dagger}(\omega)AA]\nonumber\\
&  -\ 2i\hbar K\int d\omega\lbrack\kappa_{3}A^{\dagger}A^{\dagger}a_{3}%
(\omega)-\kappa_{3}^{\ast}a_{3}^{\dagger}(\omega)AA]\nonumber\\
&  -\ i\hbar K\int d\omega\lbrack\kappa_{1}A^{\dagger}A^{\dagger}Aa_{1}%
(\omega)-\kappa_{1}^{\ast}a_{1}^{\dagger}(\omega)A^{\dagger}AA]\nonumber\\
&  -\ i\hbar K\int d\omega\lbrack\kappa_{2}A^{\dagger}A^{\dagger}Aa_{2}%
(\omega)-\kappa_{2}^{\ast}a_{2}^{\dagger}(\omega)A^{\dagger}AA]\nonumber\\
&  -\ 2i\hbar K\int d\omega\lbrack\kappa_{3}A^{\dagger}A^{\dagger}A^{\dagger
}Aa_{3}(\omega)-\kappa_{3}^{\ast}a_{3}^{\dagger}(\omega)A^{\dagger}AAA]\ .
\label{eq:b3}%
\end{align}
As was shown by Gardiner and Collett \cite{gardiner85,Gea90}, the equations of
motion Eq.~(\ref{da1/dt}) through (\ref{da3/dt}) for the baths can be
integrated to yield
\begin{equation}
\frac{1}{\sqrt{2\pi}}\int d\omega\ a_{1}(\omega)=a_{1}^{in}(t)-i\sqrt
{\frac{\pi}{2}}\kappa_{1}^{\ast}A(t)\ . \label{eq:b4}%
\end{equation}
Similarly
\begin{equation}
\frac{1}{\sqrt{2\pi}}\int d\omega\ a_{2}(\omega)=a_{2}^{in}(t)-i\sqrt
{\frac{\pi}{2}}\kappa_{2}^{\ast}A(t) \label{eq:b5}%
\end{equation}
and%
\begin{equation}
\frac{1}{\sqrt{2\pi}}\int d\omega\ a_{3}(\omega)=a_{3}^{in}(t)-i\sqrt
{\frac{\pi}{2}}\kappa_{3}^{\ast}A(t)A(t)\ . \label{eq:b6}%
\end{equation}
These expressions can be used to eliminate the $a_{n}(\omega)$ from
Eq.~(\ref{eq:b3}). Evaluating the expectation value of Eq.~(\ref{eq:b3}) with
respect to a state in which all the bath modes are in a vacuum state, that
is,
\begin{equation}
a_{n}^{in}(t)|0\rangle=0\ , \label{eq:b7}%
\end{equation}
one obtains
\begin{align}
\left\langle \frac{dH_{r}}{dt}\right\rangle  &  =-2\hbar\omega_{0}[\gamma
_{1}\langle A^{\dagger}A\rangle+\gamma_{2}\langle A^{\dagger}A\rangle
+2\gamma_{3}\langle A^{\dagger}A^{\dagger}AA\rangle]\nonumber\label{eq:b8}\\
&  -\ 2\hbar K[\gamma_{1}\langle A^{\dagger}A^{\dagger}AA\rangle+\gamma
_{2}\langle A^{\dagger}A^{\dagger}AA\rangle\nonumber\\
&  \ \ \ \ \ \ \ \ \ \ \ \ +\ 2\gamma_{3}\langle A^{\dagger}A^{\dagger
}AA\rangle+2\gamma_{3}\langle A^{\dagger}A^{\dagger}A^{\dagger}AAA\rangle
]\ .\nonumber\\
&
\end{align}
If $\omega_{0}\gg K$ then as long as the mean-field is not too large one has,
to a good approximation,
\begin{align}
\left\langle \frac{dH_{r}}{dt}\right\rangle  &  =-2\hbar\omega_{0}[\gamma
_{1}\langle A^{\dagger}A\rangle+\gamma_{2}\langle A^{\dagger}A\rangle
+2\gamma_{3}\langle A^{\dagger}A^{\dagger}AA\rangle]\nonumber\label{eq:b9}\\
&
\end{align}
This is an expression for the rate with which energy is lost from the cavity.

The power dissipated within the cavity is given by
\begin{equation}
P=\int_{0}^{L}dx\ \langle:RI^{2}:\rangle
\end{equation}
where the ": :" denotes normal ordering. Using Eq.~(\ref{R}), this can be
written as
\begin{equation}
P=\int_{0}^{l}dx\ R_{0}\langle:I^{2}:\rangle+\frac{1}{I_{c}^{2}}\int_{0}%
^{l}dx\ \Delta R\langle:I^{4}:\rangle\ .
\end{equation}
where we have allowed for the possibility that $R_{0}$ and $\Delta R$ may be
functions of the distance $x$ along the resonator, as would be the expected
case if the composition and shape of the transmission line cross section
varies with $x$. Substituting Eq.~(\ref{I(x)}) into this equation yields
\begin{align}
P  &  =\hbar\omega_{n}\int_{0}^{l}dx\ R_{0}u_{n}^{2}\langle A_{n}^{\dagger
}A_{n}\rangle\\
&  +\frac{3(\hbar\omega_{n})^{2}}{2I_{c}^{2}}\int_{0}^{l}dx\ u_{n}^{4}\Delta
R\langle A_{n}^{\dagger}A_{n}^{\dagger}A_{n}A_{n}\rangle\ .\nonumber
\end{align}
In order to make the comparison of this equation with that of Eq.~(\ref{eq:b9}%
) we consider the unloaded cavity case when the coupling through the signal
port of the cavity is set to zero, that is,
\begin{equation}
\gamma_{1}=0\ .
\end{equation}
Setting
\begin{equation}
P=-\left\langle \frac{dH_{r}}{dt}\right\rangle
\end{equation}
and keeping in mind that presently $\omega_{0}$ and $\omega_{n}$ both denote
the resonance frequency of the same selected mode, one has
\begin{equation}
\gamma_{2}=\frac{1}{2}\int_{0}^{l}dx\ u_{n}^{2}R_{0} \label{gamma_2}%
\end{equation}
and
\begin{equation}
\gamma_{3}=\frac{3\hbar\omega_{0}}{8I_{c}^{2}}\int_{0}^{l}dx\ u_{n}^{4}\Delta
R\ . \label{gamma_3}%
\end{equation}

\newpage
\bibliographystyle{plain}
\bibliography{apssamp}

\begin{thebibliography}{99}                                                                                               %


\bibitem {tucker79}J.~R.~Tucker, IEEE J. Quantum Electron. \textbf{QE-15},
1234 (1979).

\bibitem {tucker85}J.~R.~Tucker and M.~J.~Feldman, Rev. Mod. Phys.
\textbf{57}, 1055 (1985).

\bibitem {kuzmin83}L.~S.~Kuzmin, K.~K.~Likharev, V.~V.~Migulin, and
A.~B.~Zorin, IEEE Trans. Magn. \textbf{MAG-19}, 618 (1983).

\bibitem {yurke89}B.~Yurke, L.~R.~Corruccini, P.~G.~Kaminsky, L.~W.~Rupp,
A.~D.~Smith, A.~H.~Silver, R.~W.~Simon, and E.~A.~Whittaker, Phys. Rev. A,
\textbf{39}, 2519 (1989).

\bibitem {movshovich90}R.~Movshovich, B.~Yurke, P.~G.~Kaminsky, A.~D.~Smith,
A.~H.~Silver, R.~W.~Simon, and M.~V.~Schneider, Phys. Rev. Let. \textbf{65},
1419 (1990).

\bibitem {Dahm97}T. Dahm and D. J. Scalapino, J. App. Phys. \textbf{81}, 2002 (1997).

\bibitem {abdo}B. Abdo, E. Segev, O. Shtempluck, and E. Buks,
arXiv:cond-mat/0501114v2 (10 Jan 2005), arXiv:cond-mat/0504582 (22 April
2005), arXiv:cond-mat/0507056 (3 July 2005), arXiv:cond-mat/0501236 (11 Jan
2005 ).

\bibitem {villeneuve93}A. Villeneuve, C. C. Yang, G. I. Stegeman, C-H. Lin,
and H-H Lin, Appl. Phys. Lett. \textbf{62}, 2465 (1993).

\bibitem {fox95}A. M. Fox, J. J. Baumberg, M. Dabbicco, B. Huttner, and J. F.
Ryan, Phys. Rev. Lett. \textbf{74}, 1728 (1995).

\bibitem {ho95}S-T. Ho, X. Zhang, and M. K. Udo, J. Opt. Soc. Am. B
\textbf{12}, 1537 (1995).

\bibitem {zaitsev}S. Zaitsev and E. Buks, arXiv:cond-mat/053130v1 6 Mar 2005.

\bibitem {gerry93}C. C. Gerrry and E. E. Hach III, Opt. Commun. \textbf{100},
211 (1993).

\bibitem {gilles94}L.~Gilles, B.~M.~Garraway, and P.~L.~Knight, Phys. Rev. A
\textbf{49}, 2785 (1994).

\bibitem {li95}GX. Li, JS. Peng, and P. Zhou, Chinese Physics Letters
\textbf{12}, 79 (1995).

\bibitem {gardiner85}C.~W.~Gardiner and M.~J.~Collett, Phys. Rev. A
\textbf{31}, 3761 (1985).

\bibitem {Gea90}J. Gea-Banacloche, N. Lu, L. M. Pedrotti, S. Prasad, M. O.
Scully, and K. Wodkiewich, Phys. Rev. A \textbf{41}, 369 (1990).

\bibitem {imoto85}N. Imoto, H. A. Haus, and Y. Yamamoto, Phy. Rev. A
\textbf{32}, 2287 (1985).

\bibitem {white00}A. G. White, P. K. Lam, D. E. McClelland, H-A. Bachor, and
J. Munro, J. Opt. B \textbf{2}, 553 (2000).

\bibitem {yurke84}B. Yurke and J. S. Denker, Phys. Rev. A \textbf{29}, 1419 (1984).

\bibitem {tornau74}N.~Tornau and A.~Bach, Opt. Commun. \textbf{11}, 46 (1974).

\bibitem {agarwal86}G. S. Agarwal and G. P. Hildred, Opt. Comm. \textbf{58},
287 (1986).

\bibitem {gilles93}L.~Gilles and P.~L.~Knight, Phys. Rev. A \textbf{48}, 1582 (1993).

\bibitem {ezaki99}H. Ezaki, J. Phys. Soc. Japan, \textbf{68}, 1562 (1999).

\bibitem {kitamura99}M. Kitamura and T. Tokihiro, J. Opt. B \textbf{1}, 546 (1999).

\bibitem {nayfeh}A. H. Nayfeh and D. T. Mook, \textit{Nonlinear Oscillations},
(Wiley, 1979).

\bibitem {landau}L. D. Landau, \textit{Mechanics}, 3rd Ed., (Pergamon, 1976).

\bibitem {yurke95}B. Yurke, D. S. Greywall, A. N. Pargellis, and P. A. Busch,
Phys. Rev. A \textbf{51}, 4211 (1995).

\bibitem {yurke85}B. Yurke, Phys. Rev. A, \textbf{32}, 300 (1985).
\end{thebibliography}

\end{document}